# The effect of inhomogeneous phase on the critical temperature of smart meta-superconductor MgB$_2$

Honggang Chen, Yongbo Li, Guowei Chen, Longxuan Xu, Xiaopeng Zhao*

Smart Materials Laboratory, Department of Applied Physics
Northwestern Polytechnical University
Xi'an 710129, People's Republic of China
*Corresponding author: Prof. Xiaopeng Zhao，E-mail: xpzhao@nwpu.edu.cn

**Abstract**: The critical temperature ($T_C$) of $MgB_2$, one of the key factors limiting its application, is highly desired to be improved. On the basis of the meta-material structure, we prepared a smart meta-superconductor structure consisting of $MgB_2$ micro-particles and inhomogeneous phases by an ex situ process. The effect of inhomogeneous phase on the $T_C$ of smart meta-superconductor $MgB_2$ was investigated. Results showed that the onset temperature ($T_C^{on}$) of doping samples was lower than those of pure $MgB_2$. However, the offset temperature ($T_C^{off}$) of the sample doped with $Y_2O_3$:$Eu^{3+}$ nanosheets with a thickness of 2~3 nm which is much less than the coherence length of $MgB_2$ is 1.2 K higher than that of pure $MgB_2$. The effect of the applied electric field on the $T_C$ of sample was also studied. Results indicated that with the increase of current, $T_C^{on}$ is slightly increased in the samples doping with different inhomogeneous phases. When increasing current, the $T_C^{off}$ of the samples doped with nonluminous inhomogeneous phases was decreased. However, the $T_C^{off}$ of the luminescent inhomogeneous phase doping samples increased and then decreased as increasing current.

**Keywords** smart meta-superconductor $MgB_2$ • $Y_2O_3$:$Eu^{3+}$ nanosheets • Inhomogeneous phase • Applied electric field • $T_C$

## 1. Introduction

Since Akimitsu et al. [1] discovered the binary compound $MgB_2$ superconductor in 2001, this material has attracted considerable attention because of its simple structure, low cost, large coherence length, and relatively high transition temperature ($T_C$ = 39 K) [2]. Improving the superconducting transition temperature of $MgB_2$ can not only increase its application but also promote the development of superconductivity theory. The most commonly used method in improving the critical temperature ($T_C$) of $MgB_2$ is chemical doping. Substituting Mg and B with Al and C, respectively, in $MgB_2$ forms the displacement doping. However, results showed that the two kinds of doping reduce the $T_C$ of $MgB_2$ [3-7]. In addition, another possible method for improving the $T_C$ is increasing the density of holes by partially substituting Mg with Li. Nevertheless, experimental results showed that $T_C$ is still reduced [8, 9]. The $T_C$ of $MgB_2$ is reduced because of the presence of the dopant as an impurity in $MgB_2$, which results in poor grain connectivity and doping into other substances to distort the $MgB_2$ lattice. Although it is important to increase the $T_C$ of $MgB_2$ beyond the theoretical value, it also presents considerable challenge and requires further investigation.

Meta-material, a type of artificially structured composite material, is composed of the matrix material and its unit material. The properties of these materials mainly depend on the artificial structure, which can realize many special functions [10-15]. With the development of meta-material, the use of meta-material concept to design superconducting materials to improve its $T_C$ has been recognized. We proposed the introduction of electroluminescence (EL) materials in meta-materials to enhance the superconducting transition temperature through EL [16, 17]. Zhang et al. [18] used an in situ solid-phase sintering process to dope $Y_2O_3$:$Eu^{3+}$ EL materials in $MgB_2$, their results showed that the $T_C$ of $MgB_2$ can be enhanced by doping of $Y_2O_3$:$Eu^{3+}$ EL materials. However, in the in situ sintering process, the raw material B reacts with the $Y_2O_3$:$Eu^{3+}$ EL material to form the impurity phase of $YB_4$. To avoid the formation of impurity phase of $YB_4$, Tao et al. [19]

doped $Y_2O_3:Eu^{3+}$ EL material into $MgB_2$ by an ex situ solid-phase sintering process, experimental results indicated that the $Y_2O_3:Eu^{3+}$ EL material doping can improve the superconducting transition temperature of $MgB_2$, moreover, the morphology and size of the $Y_2O_3:Eu^{3+}$ EL material affect the $T_C$ of $MgB_2$. To improve the distribution and connectivity of the inhomogeneous phase dopants, the effect of different concentrations and sizes of $YVO_4:Eu^{3+}$ microsheets EL material on the superconducting transition temperature of $MgB_2$ was studied. Results showed that the $T_C^{off}$ of doped samples increases by 1.6 K compared with that of pure $MgB_2$ when the doping concentration is 2.0 wt. %. Recently, Smolyaninov et al. [20-24] proposed that material can be designed as a meta-material structure with an effective dielectric constant $\varepsilon_{eff} \approx 0$, which can improve the $T_C$ of material. They also confirmed the material performance in their experiment.

For superconducting samples, the magnitude of the test current directly affects the accuracy of the test results. Considerably low current tends to decrease the useful voltage signal of the sample. Hence, the requirement of the voltage drop measurement instrument is highly. When the current is significantly large, although the requirement of the voltage drop measurement instrument can be reduced, but it will increase the thermal effect of the sample. Consequently, a large temperature hysteresis occurs, which affects the acquisition of real data [25, 26]. Ye et al. [27, 28] found that superconducting transition in unconventional superconductors ZrNCl and $MoS_2$ through carrier doping induced by an applied electric field. Changing the applied electric field can obtain different transition temperatures, ZrNCl and $MoS_2$ display a transition temperature ($T_C$) of 15.2 and 10.8 K, respectively, on the optimum carrier doping.

On the basis of the idea of meta-material structure, our group proposed a smart meta-superconductor with a sandwich structure, where $MgB_2$ particles are used as the matrix material, the EL material distributed around the $MgB_2$ particles are as inhomogeneous phase. When evaluating the curve of the temperature dependence of resistivity ($R$–$T$) of the samples, in the local electric field, $MgB_2$ particles act as microelectrodes, which promote the EL of inhomogeneous phase EL materials, and found inhomogeneous phase significantly improves the $T_C$ [18, 19]. In this paper, the responses of the critical temperature of $MgB_2$ to inhomogeneous phase doping and changing the applied electric field are systematically studied. At first, we prepared the $Y_2O_3:Eu^{3+}$, $Y_2O_3$, and $Y_2O_3:Sm^{3+}$ nanosheets inhomogeneous phases, they were doped into $MgB_2$ by an ex situ process. The effects of doping $Y_2O_3:Sm^{3+}$/ $Y_2O_3$ nonluminous inhomogeneous phase or $Y_2O_3:Eu^{3+}$ EL inhomogeneous phase on the superconducting properties of $MgB_2$ were investigated. Then, on the basis of the theory that the material with a meta-structure and an    can improve the $T_C$ of the material, $Y_2O_3:Eu^{3+}$ microsheets and nano-Ag solution were doped into $MgB_2$ to change the    of the system, the superconducting properties of doping samples were also evaluated. Finally, we further examined the influence of the applied electric field on the EL and nonluminous inhomogeneous phases doping samples.

## 2. Experiment
### 2.1 Preparation of nanosheets / microsheets
At first, a certain amount of $Y_2O_3$ and $Eu_2O_3$ powder were added to 4 mL of concentrated nitric acid under stirring and heating at 70 °C for 1 h to obtain the $Y(Eu)(NO_3)_3$ white crystals. Afterward, some of the white crystals were dissolved in 24 mL of benzyl alcohol with constant stirring, and 6 mL of octylamine was added in the above solution and stirred for 30 min. Finally, the resulting solution was transferred to a reaction still heated at 160 °C for 24 h. The obtained precipitates were separated via centrifugation, washed several times with ethanol, and then dried in air at 60 °C for 12 h. The final products ($Y_2O_3$:$Eu^{3+}$ nanosheets, marked as N1) were prepared by calcination at 800 °C for 2 h. Moreover, $Y_2O_3$ nanosheets (marked as N2) and $Y_2O_3$:$Sm^{3+}$ nanosheets (marked as N3) were obtained by changing the raw material from the above-mentioned procedures [29], whereas $YVO_4$:$Eu^{3+}$ microsheets (marked as N4) and $Y_2O_3$:$Eu^{3+}$ microsheets (marked as N5) [30] were obtained by changing both the raw materials and the experimental conditions.

## 2.2 Preparation of doped $MgB_2$-based superconductors

The concentration of nano-Ag solution in the experiment was 2000 ppm. The particle size was 15 nm, and the solvent was anhydrous ethanol.

$MgB_2$ powder and nanosheets/microsheets dopants ($MgB_2$ powder, microsheets dopant, and nano-Ag solution) were mixed in 15 mL of ethanol to form a suspension. The suspension was transferred into a culture dish after 30 min of sonication. Subsequently, the dish was placed in a vacuum oven for 4 h at 60 °C. The resultant black-powder was pressed into tablets. Finally, the tablets were placed in tantalum vessels and annealed at 800 °C for 2 h at heating and cooling rates were of 10 °C min$^{-1}$ and 5 °C min$^{-1}$, respectively. Afterward, the final products were obtained. For convenience of description, symbols were used to represent the samples. Furthermore, a pure $MgB_2$ sample marked as A was prepared for comparison. The symbols, dopant types, and dopant concentrations of each sample are shown in Table 1.

**Table 1** Symbols, dopant types, and dopant concentrations of each sample.

| Symbols of sample | A | B | C | D | E | F | G | H |
|---|---|---|---|---|---|---|---|---|
| Dopant types | none | N2 | N3 | N1 | N4 | N5 | N5 | N5 |
| Dopant concentrations (wt. %) | 0 | 2 | 2 | 2 | 2 | 2 | 2 | 2 |
| Nano-Ag concentrations (wt. %) | 0 | 0 | 0 | 0 | 0 | 0 | 0.2 | 0.4 |

## 3. Results and discussion

Figs. 1a, 1b, and 1c shows the SEM images of $YVO_4$:$Eu^{3+}$ microsheets and $Y_2O_3$:$Eu^{3+}$ microsheets, and the TEM image of $Y_2O_3$:$Eu^{3+}$ nanosheets respectively. 1d shows the AFM image of the $Y_2O_3$:$Eu^{3+}$ nanosheets. 1e present the thickness of $Y_2O_3$:$Eu^{3+}$ nanosheets. We know the prepared of $YVO_4$:$Eu^{3+}$ and $Y_2O_3$:$Eu^{3+}$ sheets show varying sizes. The sizes of $YVO_4$:$Eu^{3+}$ and $Y_2O_3$:$Eu^{3+}$ microsheets are 1–2 and 0.5–1 μm, respectively, and the size of $Y_2O_3$:$Eu^{3+}$ nanosheets are approximately 50 nm. The thickness of the $Y_2O_3$:$Eu^{3+}$ nanosheets are about 2~3 nm, which is much less than the coherence length of $MgB_2$.

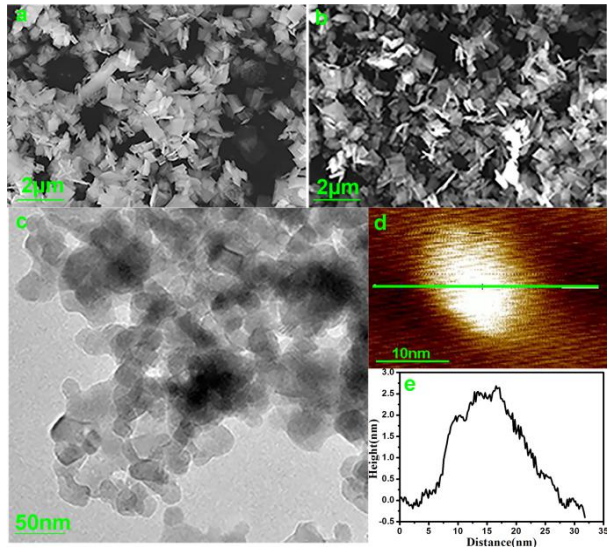

**Fig. 1** SEM images of (a) YVO$_4$:Eu$^{3+}$ microsheets (N4), (b) Y$_2$O$_3$:Eu$^{3+}$ microsheets (N5) and (c) TEM image of Y$_2$O$_3$:Eu$^{3+}$ nanosheets (N1); (d) AFM image of the Y$_2$O$_3$:Eu$^{3+}$ nanosheets. (e) Height profile corresponding to the lines draw in d.

Fig. 2 shows the X–ray diffraction (XRD) patterns of the partial samples and the SEM image of pure MgB$_2$. XRD results showed that the main phase is MgB$_2$. A small number of MgO and Mg impurities are also observed, which may broaden the superconducting transition temperature. MgO, which was formed during the preparation of MgB$_2$, is present in all samples. We also observed the existence of the Y$_2$O$_3$ in the doping samples. The SEM image shows that the irregularly shaped MgB$_2$ particles are mainly about 0.2-2 μm in size. The boundary between particles is also evident, which also broaden the superconducting transition temperature.

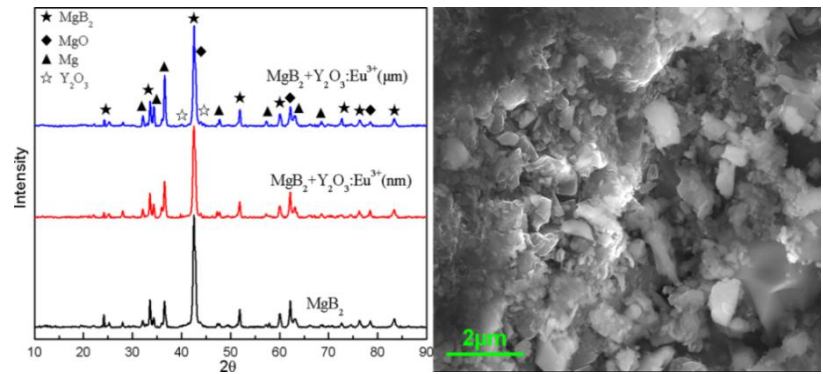

**Fig. 2** X–ray diffraction patterns of MgB$_2$ (A), MgB$_2$+2 wt. % Y$_2$O$_3$:Eu$^{3+}$ (nm) (D), MgB$_2$+2 wt. % Y$_2$O$_3$:Eu$^{3+}$ (μm) (F), and SEM image of pure MgB$_2$ (A)

The XRD pattern of MgB$_2$ doped with Y$_2$O$_3$:Eu$^{3+}$ nanosheets is shown in Fig. 2. Since the low content of Y$_2$O$_3$:Eu$^{3+}$ nanosheets, the Y$_2$O$_3$ peak is not obvious in the XRD pattern. To further prove that the sample is adulterated with Y$_2$O$_3$:Eu$^{3+}$ nanosheets, we performed an elemental analysis of the sample, and results are shown in Fig. 3. Fig. 3a presents the SEM image of MgB$_2$ doped with Y$_2$O$_3$:Eu$^{3+}$ nanosheets; Figs. 3b-3d illustrate the distribution of certain chemical elements, the corresponding element is listed at the top right corner of each figure. According to the element distribution map, the sample contains a large number of Mg, and $Y_2O_3$:$Eu^{3+}$ nanosheets inhomogeneous phase dopants distributed around the $MgB_2$ particles.

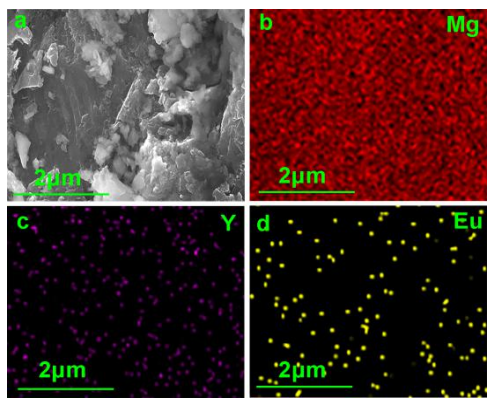

**Fig. 3** SEM image of (a) $MgB_2$+2 wt. % $Y_2O_3$:$Eu^{3+}$ (nm) (D), and chemical element distribution map (b-d)

Fig. 4 depicts the $R$–$T$ curve of the pure $MgB_2$, and $MgB_2$ doped with $Y_2O_3$:$Eu^{3+}$, $Y_2O_3$ and $Y_2O_3$:$Sm^{3+}$ nanosheets. The two characteristic temperatures, namely, $T_C^{on}$ and $T_C^{off}$, on each $R$–$T$ curve are discussed. (1) The black curve shows the $R$–$T$ curve of pure $MgB_2$ (A). The $T_C^{on}$ and $T_C^{off}$ of pure $MgB_2$ are 38.2 K and 33.6 K, respectively, the range of superconducting transition temperature is broad, this phenomenon is largely attributed to that the samples contain MgO and Mg impurities and exhibit poor grain connectivity [31]. (2) The resistivity of $MgB_2$ doped with nanosheets is higher than that of pure $MgB_2$. (3) The $T_C^{off}$ of $MgB_2$ doped with $Y_2O_3$:$Eu^{3+}$ nanosheets increases by 1.2 K compared with that of pure $MgB_2$, which may be due to the $Y_2O_3$:$Eu^{3+}$ nanosheets EL material distributed around the $MgB_2$ particles to form a special response meta-structure. $Y_2O_3$:$Eu^{3+}$ nanosheets would generate an EL during the measurement of $R$–$T$ curve of the sample, which may improve the superconducting transition temperature [19]. (4) The $T_C^{off}$ of $MgB_2$ doped with $Y_2O_3$ and $Y_2O_3$:$Sm^{3+}$ nanosheets are lower than pure $MgB_2$ sample. The $MgB_2$ doped with $Y_2O_3$:$Sm^{3+}$ nanosheets presents the lowest $T_C^{off}$. The $T_C^{on}$ of all of the $MgB_2$ doped with nanosheets is lower than pure $MgB_2$. The $T_C^{on}$ values of $MgB_2$ doped with $Y_2O_3$ and $Y_2O_3$:$Eu^{3+}$ nanosheets are reduced by 0.2 K, and $MgB_2$ doped with $Y_2O_3$:$Sm^{3+}$ nanosheets is reduced by 0.4 K.

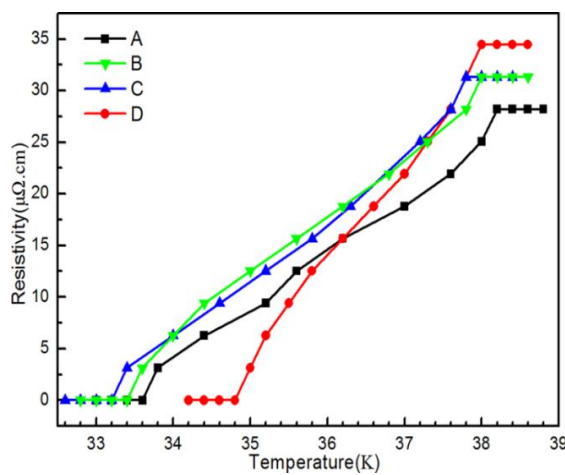

**Fig. 4** Temperature-dependent resistivity of $MgB_2$ doped with different Nano-sheets (pure $MgB_2$ (A), $MgB_2$+2 wt. % $Y_2O_3$ (nm) (B), $MgB_2$+2 wt. % $Y_2O_3$:$Sm^{3+}$ (nm) (C), $MgB_2$+2 wt. % $Y_2O_3$:$Eu^{3+}$ (nm) (D))

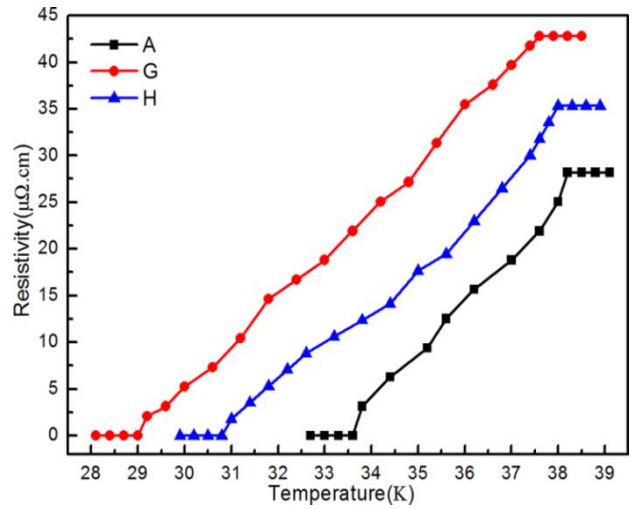

**Fig. 5** Temperature-dependent resistivity of $MgB_2$ doped with $Y_2O_3$:$Eu^{3+}$ micro sheets and Nano-Ag solution ($MgB_2$ (A), $MgB_2$+2 wt. % $Y_2O_3$:$Eu^{3+}$ (μm) +0.2 wt. % Ag (G), $MgB_2$+2 wt. % $Y_2O_3$:$Eu^{3+}$ (μm) +0.4 wt. % Ag (H))

    Smolyaninov et al. proposed that $MgB_2$ doped with 5 nm diamond particles can make the dielectric constant close to 0; consequently, the superconducting transition temperature of the $MgB_2$-based metamaterial superconductor can reach the liquid nitrogen temperature [32]. On the basis of the idea that changing the dielectric constant can increase the superconducting transition temperature of meta-superconductor, we prepared $MgB_2$ doped with $Y_2O_3$:$Eu^{3+}$ microsheets and nano-Ag solution to change the of the system so as to improve the superconducting transition temperature of $MgB_2$. Fig. 5 presents the *R–T* curve of the $MgB_2$ doped with $Y_2O_3$:$Eu^{3+}$ microsheets and nano-Ag solution. However, this graph indicated that the $MgB_2$ doped with $Y_2O_3$:$Eu^{3+}$ microsheets and nano-Ag solution fails to improve the $T_C$ of $MgB_2$. All of the doped samples exhibit superconducting transition. The resistivity in the normal state increases, and the superconducting transition temperature of $MgB_2$ doped with nano-Ag solution decreases remarkably.

**Table 2** The $T_C^{eff}$ and $T_C^{p,n}$ of doped samples under different currents.

|        | A/K       | B/K       | C/K       | D/K       | E/K       | F/K       |
|--------|-----------|-----------|-----------|-----------|-----------|-----------|
| 50 mA  | 33.8-38.2 | 33.6-37.6 | 33.4-37.6 | 34.6-37.8 | 34.6-37.8 | 34.2-38   |
| 100 mA | 33.6-38.2 | 33.4-38   | 33.2-37.8 | 34.8-38   | 35.2-38   | 34.6-38   |
| 200 mA | 33.4-38.2 | 33-38     | 33-38     | 34.4-38.2 | 34.8-38   | 34.4-38   |
| 300 mA | 33.2-38.2 | 32.8-38   | 32.8-38   | 34.2-38.4 | 34.6-38   | 34.2-38.2 |
| 500 mA | 33-38.2   | 32.6-38   | 32.6-38   | 34-38.4   | 34.2-38.2 | 34-38.4   |

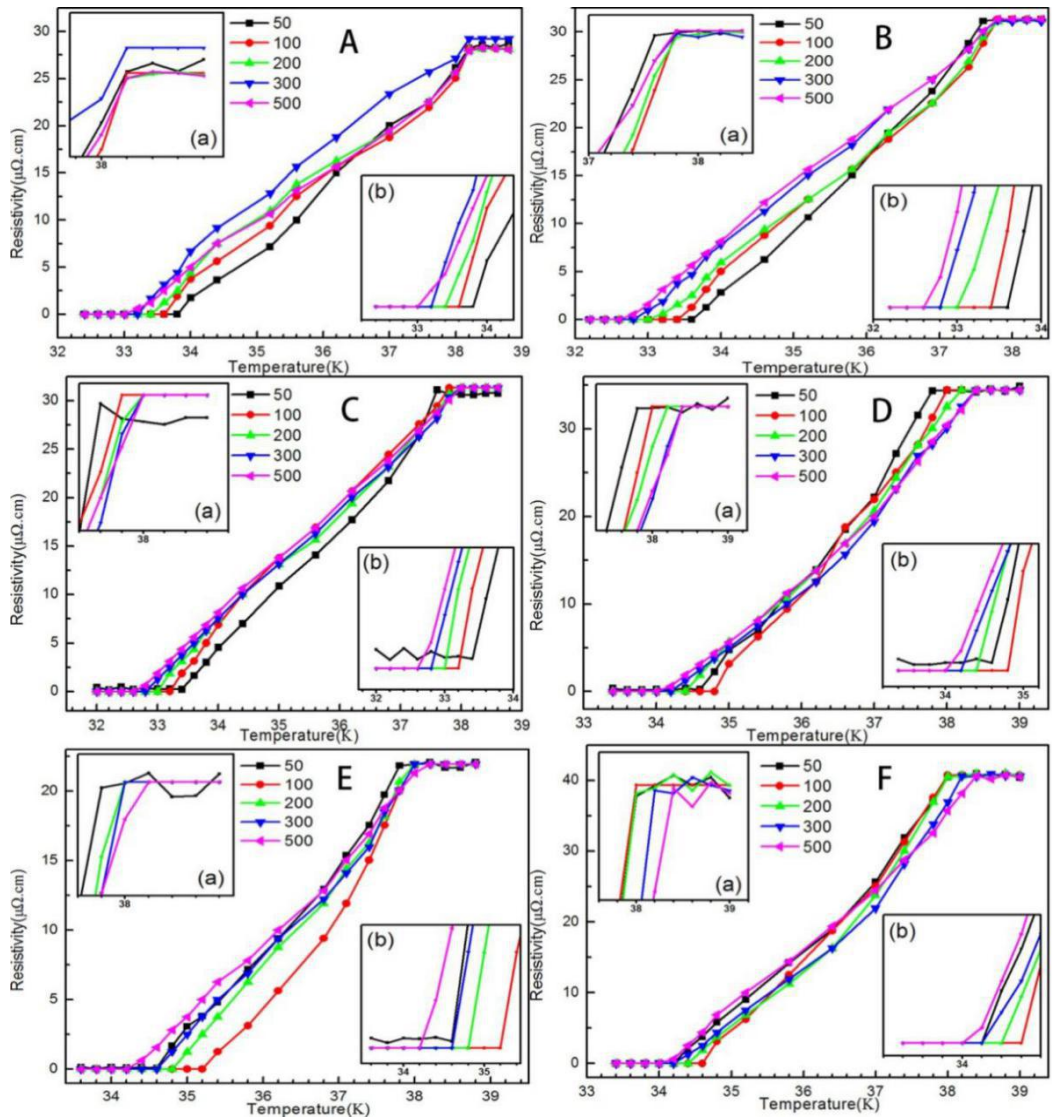

**Fig. 6** Temperature-dependent resistivity of doped samples under different currents (50, 100, 200, 300, 500 mA)

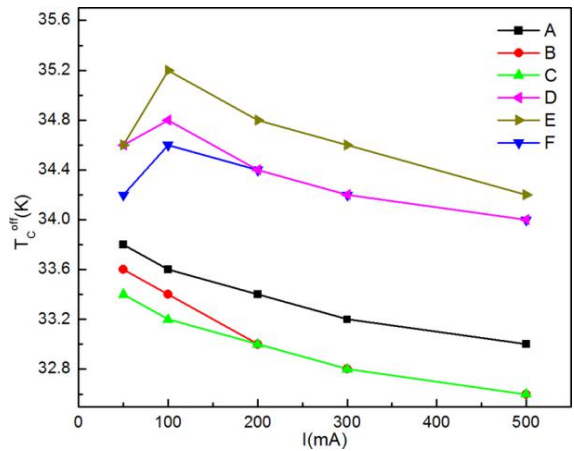

**Fig. 7** The relationship between the $T_C^{off}$ of doped samples and currents

Fig. 6 presents the *R–T* curves of six different doped samples under different currents, the upper left and bottom right corners are enlarged graphs of corresponding $T_C^{on}$ and $T_C^{off}$, respectively. Table 2 provides a list of the concrete values of $T_C^{on}$ and $T_C^{off}$. Fig. 7 shows the relationship between the $T_C^{off}$ of doped samples and currents. According to the results presented in Fig. 6, Fig. 7, and Table 2, the following trends are determined:(1) With increasing current, the $T_C^{off}$ values of pure MgB$_2$ (A), and MgB$_2$ doped with Y$_2$O$_3$ nanosheets (B) and Y$_2$O$_3$:Sm$^{3+}$ nanosheets (C) decrease, and the $T_C^{on}$ and $T_C^{off}$ of MgB$_2$ doped with Y$_2$O$_3$ nanosheets and Y$_2$O$_3$:Sm$^{3+}$ nanosheets are lower than those of pure MgB$_2$ at corresponding current, this result is attributed to impurity doping. (2) The $T_C^{off}$ of MgB$_2$ doped with Y$_2$O$_3$:Eu$^{3+}$ micro-sheets (F), Y$_2$O$_3$:Eu$^{3+}$ nanosheets (D), and YVO$_4$:Eu$^{3+}$ micro-sheets (E) show a slight increase at a current less than or equal to 100 mA. However, when the current lager than 100 mA, $T_C^{off}$ decrease with increasing current. This result may be due to the considerably high current, which results in a remarkable thermal effect. Consequently, the temperature gradient in the superconducting sample is increased, thereby increasing the influence of thermoelectric potential on the measurements. (3) The $T_C^{on}$ of the pure MgB$_2$ sample remains unchanged when increasing current, whereas that of the doped samples show a slight increase. The $T_C^{on}$ values of all nonluminous inhomogeneous phases doping samples are no more than 38.2 K, but those of samples doped with EL inhomogeneous phases are more than 38.2 K at a certain current. Japanese scientists observed superconducting transition by changing the electric field in unconventional superconductors ZrNCl and MoS$_2$. ZrNCl and MoS$_2$ show $T_C$ of 15.2 and 10.8 K, respectively, on the optimum carrier doping [27, 28]. Nevertheless, in our experiment the effect of changing electric field on the $T_C$ is not obvious, this may be due to we change the carrier density by changing the current directly rather than changing electric field to induce carrier density change.

## 4. Conclusion

On the basis of the smart metamaterial superconductor model, we found that inhomogeneous phase significantly improves the $T_C$ of superconductor. In this paper, the responses of the $T_C$ of MgB$_2$ to inhomogeneous phase doping and changing applied electric field are systematically investigated. At first, we prepared Y$_2$O$_3$:Eu$^{3+}$ and Y$_2$O$_3$, Y$_2$O$_3$:Sm$^{3+}$ nanosheets inhomogeneous phases, which were doped into MgB$_2$ by an ex situ process. Results showed that the $T_C^{off}$ of MgB$_2$ doped with Y$_2$O$_3$:Eu$^{3+}$ nanosheets increase by 1.2K compared with that of pure MgB$_2$, whereas the $T_C^{off}$ of MgB$_2$ doped with Y$_2$O$_3$ and Y$_2$O$_3$:Sm$^{3+}$ nanosheets decrease, and the MgB$_2$ doped with Y$_2$O$_3$:Sm$^{3+}$ nanosheets presents the lowest $T_C^{off}$. In addition, the $T_C^{on}$ of all of the MgB$_2$ doped with nanosheets decrease. The distribution of certain chemical elements reveals that Y$_2$O$_3$:Eu$^{3+}$ nanosheets inhomogeneous phase dopants distributed around the MgB$_2$ particles and form a meta-structure. Hence, the effectiveness of Y$_2$O$_3$:Eu$^{3+}$ nanosheets improve the $T_C$ of MgB$_2$ can be fully reflected. Then, on the basis of the idea that changing the dielectric constant can increase the superconducting transition temperature of meta-superconductor, we prepared MgB$_2$ doped with Y$_2$O$_3$:Eu$^{3+}$ microsheets and nano-Ag solution to change the   of the system so as to improve the superconducting transition temperature of MgB$_2$. Nevertheless, experimental results show that the codoping can not improve the $T_C$ of MgB$_2$. Additionally, the superconducting transition temperature of MgB$_2$ doped with nano-Ag solution decrease remarkably, and the resistivity in the normal state increase. Finally, we also find that the applied electric field affects the $T_C$ of doping samples, when increasing test current, the $T_C^{off}$ of nonluminous inhomogeneous phase doping samples decrease. However, the $T_C^{off}$ of luminescent inhomogeneous phase doping samples increase and then decrease. The $T_C^n$ of pure $MgB_2$ showed no change, whereas the $T_C^n$ of doped samples can more than 38.2 K at certain conditions. Improving the superconducting transition temperature of $MgB_2$ can not only increase its application but also promote the development of superconductivity theory. This study provides a further exploration for the considerable challenge of improving the $T_C$ of smart meta-superconductor $MgB_2$.

**Acknowledgments** This work was supported by the National Natural Science Foundation of China for Distinguished Young Scholar under Grant No. 50025207.